# Synchronization of Fractional-order Chaotic Systems with Gaussian fluctuation by Sliding Mode Control

Yong Xu[*], Hua Wang

Department of Applied Mathematics

Northwestern Polytechnical University, Xi'an, 710072, China

**Abstract**

This paper is devoted to the problem of synchronization between fractional-order chaotic systems with Gaussian fluctuation by the method of fractional-order sliding mode control. A fractional integral (FI) sliding surface is proposed for synchronizing the uncertain fractional-order system, and then the sliding mode control technique is carried out to realize the synchronization of the given systems. One theorem about sliding mode controller is presented to prove the proposed controller can make the system synchronize. As a case study, the presented method is applied to the fractional-order Chen-Lü system as the drive-response dynamical system. Simulation results show a good performance of the proposed control approach in synchronizing the chaotic systems in presence of Gaussian noise.

**Keywords**

Synchronization, fractional-order, chaotic system, Gaussian fluctuation, sliding mode control

## 1. Introduction

Synchronization, which means "things occur at the same time or operate in unison", has received a great deal of interest among scientists from various fields in the last few years [1-4], especially in fractional-order chaotic systems. It has been recognized that many systems in interdisciplinary fields can be elegantly described by fractional-order differential equations, such as viscoelastic materials [5], electrical circuits [6], population models [7], financial systems [8], and so on. Meanwhile, most of precious studies have been shown that some fractional-order systems exhibit chaotic behavior [9-14]. In particular, fractional-order chaotic behavior has wide promising applications in information encryption, image processing, secure communication, etc [15-18]. Therefore, synchronization of fractional chaotic systems starts to attract increasing attentions due to its great importance in applications ranging from physics to engineering, from computer science to biology, and even from economics to brain science.

[*] Corresponding author E-mail: hsux3@nwpu.edu.cn
Tel./fax :86-29-88431637

A basic configuration for chaos synchronization is the drive-response pattern, where the response of chaotic system must track the drive chaotic trajectory. Some approaches based on this configuration have been attained to achieve chaos synchronization in fractional-order chaotic systems, such as PC control [19], active control [20], adaptive control [21,22], sliding mode control (SMC) [23], and a scalar transmitted signal method [24], etc. In which, the sliding mode controller has some attractive advantages, including: (i) fast dynamic responses and good transient performance; (ii) external disturbance rejection; and (iii) insensitivity to parameter variations and model uncertainties [25-26]. In addition, SMC method plays an important role in the application to practical problems. For example, in [27], Tavazoei MS and his co-operator proposed a controller based on active sliding mode theory to synchronize chaotic fractional-order systems in master-slave structure. In [2], the problem of modified projective synchronization of fractional-order chaotic system was considered, and finite-time synchronization of non-autonomous fractional-order uncertain chaotic systems was investigated by Aghababa MP in [28].

All of the methods mentioned above have been used to synchronize the deterministic fractional-order chaotic systems. However, the synchronization of two different chaotic systems in uncertain environment plays a significant role in practical applications, and this problem will be more challenging and difficult if the fractional-order dynamic systems are influenced by some uncertainty. In this investigation, our aim is to synchronize two fractional-order chaotic systems with uncertain environment. To achieve this goal, we have proposed fractional integral (FI) sliding mode surface which combines the property of fractional-order equation with sliding mode control method. One theorem about sliding mode controller is presented and proved that the proposed controller can make the system synchronize well. Then a numerical example is given to verify the effectiveness of the mentioned method, and the good agreements are also found between the theoretical and the numerical results.

This paper is organized as follows. In section 2, two fractional-order chaotic systems with uncertainty and problem formulation are presented. In section 3, we investigate the design method of sliding mode controller, and one theorem is obtained to prove the effectiveness of proposed controller. One example is presented to carry out the numerical simulations in section 4. Finally, conclusions are presented to close this paper.

## 2. System description and problem formulation

Consider the following class of fractional-order chaotic system with noise, which is described by

$$D^\alpha \boldsymbol{x} = \boldsymbol{A}_1 \boldsymbol{x} + \boldsymbol{f}_1(\boldsymbol{x}) + \boldsymbol{h}(\boldsymbol{x},t)\boldsymbol{W}(t), \quad (1)$$

where $\boldsymbol{x} = [x_1, x_2, ..., x_n]^T \in R^n$ denotes the state vector, $\boldsymbol{A}_1 \in R^{n \times n}$ is a constant matrix, $\boldsymbol{f}_1 : R^n \to R^n$ is nonlinear vector function, $\boldsymbol{h}$ is noise intensity function which is sufficient smooth and bounded, i.e. $|\boldsymbol{h}(\boldsymbol{x},t)| \leq H$ ($H$ is a positive number), $\boldsymbol{W}(t) = [W_1(t), W_2(t), ..., W_n(t)]^T$ is $n$-dimensional Brownian motion. Accordingly, $\dot{\boldsymbol{W}}(t) = [\dot{W}_1, \dot{W}_2, ..., \dot{W}_n]^T$ is a $n$-dimensional Gaussian white noise vector, in which every two noises are statistical independent. And $D^\alpha$ is the Caputo derivative, $\alpha = [\alpha_1, \alpha_2, ..., \alpha_n]^T$ is the order of fractional derivative, which is defined as

$$D^\alpha f(t) = \frac{1}{\Gamma(n-\alpha)} \int_0^t \frac{f^n(\tau)}{(t-\tau)^{\alpha-n+1}} d\tau, \text{ for } n-1 < \alpha < n \quad (2)$$

with $\Gamma(z) = \int_0^\infty e^{-t} t^{z-1} dt$ the Euler's Gamma function.

Let system (1) be the driving system, then response system with a controller

$u(t) = [u_1(t), u_2(t), ..., u_n(t)]^T$ is given by

$$D^\alpha y = A_2 y + f_2(y) + u(t), \quad (3)$$

where $\alpha = [\alpha_1, \alpha_2, ..., \alpha_n]^T$ is the order of fractional derivative, $y = [y_1, y_2, ..., y_n]^T \in R^n$ is state vector, $A_2 \in R^{n \times n}$ is a coefficient matrix, $f_2 : R^n \to R^n$ is a function matrix. Let $e = [e_1, e_2, ..., e_n]^T = y - x$. Then, from system (1) and (3), one has the error dynamics

$$\begin{aligned} D^\alpha e &= A_2 y + f_2(y) + u(t) - A_1 x - f_1(x) - h(x,t)\dot{W} \\ &= A_2 e + F(x, y) - h(x,t)\dot{W} + u(t), \end{aligned} \quad (4)$$

where $F(x, y) = f_2(y) - f_1(x) + (A_2 - A_1)x$.

Thus, the control problem considered in this study is that for chaotic driving system (1) and response system (3), the two fractional order systems are to be synchronized by designing an appropriate controller $u(t)$ such that

$$\lim_{t \to \infty} \|e\| = \lim_{t \to \infty} \|y - x\| = 0, \quad (5)$$

where $\|.\|$ is defined as $\|e(t)\| = (E[e(t)^T e(t)])^{\frac{1}{2}}$, where $E[.]$ is the expected value function.

### 4. Sliding mode controller design and analysis

In the following context, we shall design sliding mode controller to establish the synchronization between driving system (1) and response system (3).

*4.1 Sliding mode controller design process*

Now, the control input vector $u(t)$ is defined to eliminate the nonlinear part of the error dynamics:

$$u(t) = H(t) - F(x, y), \quad (6)$$

where $H(t) = Kw(t)$, $K$ is a constant gain matrix. $w(t) \in R$ is the control input that satisfies

$$w(t) = \begin{cases} w^+(t) & S(e) \geq 0, \\ w^-(t) & S(e) < 0, \end{cases}$$

in which $S = S(e)$ is a switching surface to prescribe the desired sliding mode dynamics.

So the error system (4) is then rewritten as

$$D^\alpha e = A_2 e + u(t) - h(x,t)\dot{W}. \tag{7}$$

Here, a new fractional integral (FI) switching surface is given as follows:

$$S = D^{\alpha-1} e - \int_0^t (K + A_2) e(\tau) d\tau, \tag{8}$$

where $S = [s_1, s_2, ..., s_n]^T$, and $K + A_2$ should be stable, namely, the eigenvalues $\lambda_i (i = 1, 2, ..., n)$ of matrix $K + A_2$ are negative $(\lambda_i < 0)$.

As we all know, when the system is controllable in the sliding mode, the switching surface should satisfy the following conditions:

$$S(e) = 0, \tag{9a}$$

together with:

$$\dot{S}(e) = 0. \tag{9b}$$

Substitute equations (7) and (8) into (9b), one obtains

$$\begin{aligned}
\dot{S} &= D^\alpha e - (K + A_2) e(t) \\
&= A_2 e + kw(t) - h(x,t)\dot{W} - (K + A_2) e(t) \\
&= Kw(t) - h(x,t)\dot{W} - Ke(t) \\
&= 0.
\end{aligned} \tag{10}$$

Therefore, the equivalent control law is obtained by

$$w_{eq}(t) = e(t) + K^{-1} h(x,t)\dot{W}. \tag{11}$$

In real-world applications, the Gaussian white noise $\dot{W}$ is uncertain. Therefore, the equivalent control input is modified to

$$w_{eq}(t) = e(t), \tag{12}$$

To design the sliding mode controller, we consider the constant plus proportional rate reaching law [29-30] that is

$$\dot{S} = -rS - \rho \operatorname{sgn}(S), \tag{13}$$

where $\text{sgn}(S) = [\text{sgn}(s_1), \text{sgn}(s_2), ..., \text{sgn}(s_n)]^T$, $r$ and $\rho$ are all positive numbers, $\text{sgn}(x)$ represents sign function, that is

$$\text{sgn}(x) = \begin{cases} 1, & x > 0, \\ 0, & x = 0, \\ -1, & x < 0. \end{cases}$$

So, we can get the controller

$$w(t) = K^{-1}(Ke - rS - \rho \text{sgn}(S)). \tag{14}$$

Further, according to the control law and the updated law the controller are given by

$$u(t) = Ke - rS - \rho \text{sgn}(S) - f_2(y) + f_1(x) - (A_2 - A_1)x. \tag{15}$$

And the error system can be rewritten in the following differential form:

$$D^\alpha e = (K + A_2)e(t) - h(x,t)\dot{W} - rS - \rho \text{sgn}(S). \tag{16}$$

*4.2 Synchronization analysis*

**Theorem 1** If the controller is selected as equation (15), with suitably selected $r$ and $\rho$, then the synchronization of fractional-order chaotic systems between driving system (1) and response system (3) can be achieved (i.e., the synchronization error converge to zero in the mean square norm).

*Proof* Consider a Lyapunov function constructed by the mean square norm of $S(t)$ and its differential form:

$$V = \frac{1}{2}\|S(t)\|^2 = \frac{1}{2}E[S^2(t)], \tag{17}$$

$$dV = \frac{1}{2}E[d(S^2(t))]. \tag{18}$$

According to the definition of derivative, it is obtained that:

$$\begin{aligned} d(S^2(t)) &= (S(t) + dS(t))^2 - S^2(t) \\ &= 2S(t)dS(t) + dS(t)dS(t). \end{aligned} \tag{19}$$

While, from equations (10) and (14), one gets

$$dS(t) = (-rS - \rho \text{sgn}(S))dt - h(x,t)dW. \tag{20}$$

Substituting equation (20) into equation (19) result in:

$$d(S^2(t)) = 2S(t)((-rS - \rho \operatorname{sgn}(S))dt - h(x,t)dW) + h^2(x,t)dt, \quad (21)$$

Taking expectations to equation (21) and using the properties of Brownian motion mentioned, we have

$$dV = E\left[S(t)((-rS - \rho \operatorname{sgn}(S))dt - h(x,t)dW) + \frac{1}{2}h^2(x,t)dt\right]$$
$$= E\left[S(t)((-rS - \rho \operatorname{sgn}(S))dt) + \frac{1}{2}h^2(x,t)dt\right], \quad (22)$$

Therefore:

$$\dot{V} = E\left[S(t)(-rS - \rho \operatorname{sgn}(S)) + \frac{1}{2}h^2(x,t)\right]$$
$$\leq -rE\left[S^2\right] - \rho E\left[|S|\right] + \frac{1}{2}H^2. \quad (23)$$

Equation (23) implies that as long as suitable $r$ and $\rho$ which satisfies $\frac{1}{2}H^2 \leq rE\left[S^2\right] + \rho E\left[|S|\right]$ is selected, namely, $\dot{V} \leq 0$, according to barbalat's lemma [31], system (1) and system (3) can be achieved synchronization under the controller law in (15).

This completes the proof. □

**5. Simulation**

In this part, to confirm the validity of proposed method, we numerically examine the synchronization between fractional-order Chen and fractional-order Lü systems. In the simulation, step-by-step method is performed to receive numerical solution, and the detailed descriptions of this algorithm are available in [32-33].

Here, we assume that the Chen system drives the Lü system. Hence, the driving system (fractional-order Lü system) is described as

$$\begin{cases} D^\alpha x_1 = a_1(x_2 - x_1) + h_1(x,t)\dot{W}_1, \\ D^\alpha x_2 = -x_1 x_3 + c_1 x_2 + h_2(x,t)\dot{W}_2, \\ D^\alpha x_3 = x_1 x_2 - b_1 x_3 + h_3(x,t)\dot{W}_3. \end{cases} \quad (24)$$

which can also be written as

$$D^\alpha x = A_1 x + f_1(x) + h(t, x)\dot{W}, \tag{25}$$

here,

$$x = \begin{pmatrix} x_1 \\ x_2 \\ x_3 \end{pmatrix}, A_1 = \begin{bmatrix} -a_1 & a_1 & 0 \\ 0 & c_1 & 0 \\ 0 & 0 & -b_1 \end{bmatrix}, f_1(x) = \begin{pmatrix} 0 \\ -x_1 x_3 \\ x_1 x_2 \end{pmatrix}, h(t, x) = \begin{pmatrix} h_1(t, x) \\ h_2(t, x) \\ h_3(t, x) \end{pmatrix}, \dot{W} = \begin{pmatrix} \dot{W}_1 \\ \dot{W}_2 \\ \dot{W}_3 \end{pmatrix},$$

which has been shown that the fractional-order Lü system can demonstrate chaotic behavior [25] when $a_1 = 35, b_1 = 3, c_1 = 28, \alpha = 0.9$.

The response system (fractional-order Chen system) is given as follows:

$$\begin{cases} D^\alpha y_1 = a_2(y_2 - y_1), \\ D^\alpha y_2 = (c_2 - a_2)y_1 - y_1 y_3 + c_2 y_2, \\ D^\alpha y_3 = y_1 y_2 - b_2 y_3. \end{cases} \tag{26}$$

which can be written in the following form

$$D^\alpha y = A_2 y + f_2(y), \tag{27}$$

where $y = \begin{pmatrix} y_1 \\ y_2 \\ y_3 \end{pmatrix}, A_2 = \begin{bmatrix} -a_2 & a_2 & 0 \\ c_2 - a_2 & c_2 & 0 \\ 0 & 0 & -b_2 \end{bmatrix}, f_2(y) = \begin{pmatrix} 0 \\ -y_1 y_3 \\ y_1 y_2 \end{pmatrix}.$

In which, system will exhibit chaotic behavior [25] when $a_2 = 35, b_2 = 3, c_2 = 28$, $\alpha = 0.9$. According to (3), the response system with controller can be described as follows:

$$D^\alpha y = A_2 y + f_2(y) + u(t), \tag{28}$$

where $u(t) = [u_1, u_2, u_3]^T$ is the control vector.

Now we apply the proposed sliding control approach to finish synchronization between fractional-order Lü system driven by Gaussian white noise and fractional-order Chen system. Here, we define the error states as

$$e_i = y_i - x_i, \tag{29}$$

and the sliding mode surface as

$$S = D^{\alpha-1} e - \int_0^t (K + A_2) e(\tau) d\tau, \tag{30}$$

The control law is given by

$$u(t) = Ke - rS - \rho \operatorname{sgn}(S) - f_2(y) + f_1(x) - (A_2 - A_1)x. \tag{31}$$

In the numerical simulations, the initial conditions are set as $(x_1(0), x_2(0), x_3(0)) = (7, -4, 4)$, $(y_1(0), y_2(0), y_3(0)) = (1, 3, -1)$. The noise intensity matrices are presumably given in the form of $(h_1(t,x), h_2(t,x), h_3(t,x)) = (0.3, 0.4, \sin t)$. In fact, controller parameters can be chosen as $r = 5$, $\rho = 0.5$, $K = \begin{bmatrix} 34 & -35 & 0 \\ 7 & -29 & 0 \\ 0 & 0 & 2 \end{bmatrix}$. The time step size employed in this simulation is $h = 0.0005$. Then the simulation results are summarized in Fig.1-Fig.4. The state trajectories of the system (25) and system (28) under the sliding mode control method are shown in Fig. 1(a) (signals $x_1$; $y_1$), Fig. 1(b) (signals $x_2$; $y_2$), Fig. 1(c) (signals $x_3$; $y_3$), respectively. Note that the driving system is shown by red lines where as response system is shown by blue lines. As one can see, the designed controller is effectively capable to achieve the synchronization of fractional-order Chen chaotic system, that is the state variables $(y_1, y_2, y_3)$ follow the trail of $(x_1, x_2, x_3)$ well. Then the synchronization errors between the uncertain fractional-order chaotic Lü system and fractional-order chaotic Chen system are depicted in Fig. 2(a) (signal $e_1$), Fig. 2(b) (signal $e_2$), Fig. 2(c) (signal $e_3$). As it is expected, the synchronization errors (33) close to zero. Further, the expectation and variance of error vectors $e_1$, $e_2$ and $e_3$ converge to zero, as displayed in Fig. 3 and Fig.4, which all indicate that the chaos synchronization between uncertain fractional-order chaotic Lü and Chen systems are indeed realized.

From the simulation results, it can be concluded that the obtained theoretic results are efficient and feasible for synchronizing fractional uncertain dynamical systems and the proposed controller guarantees the convergence of the error system.

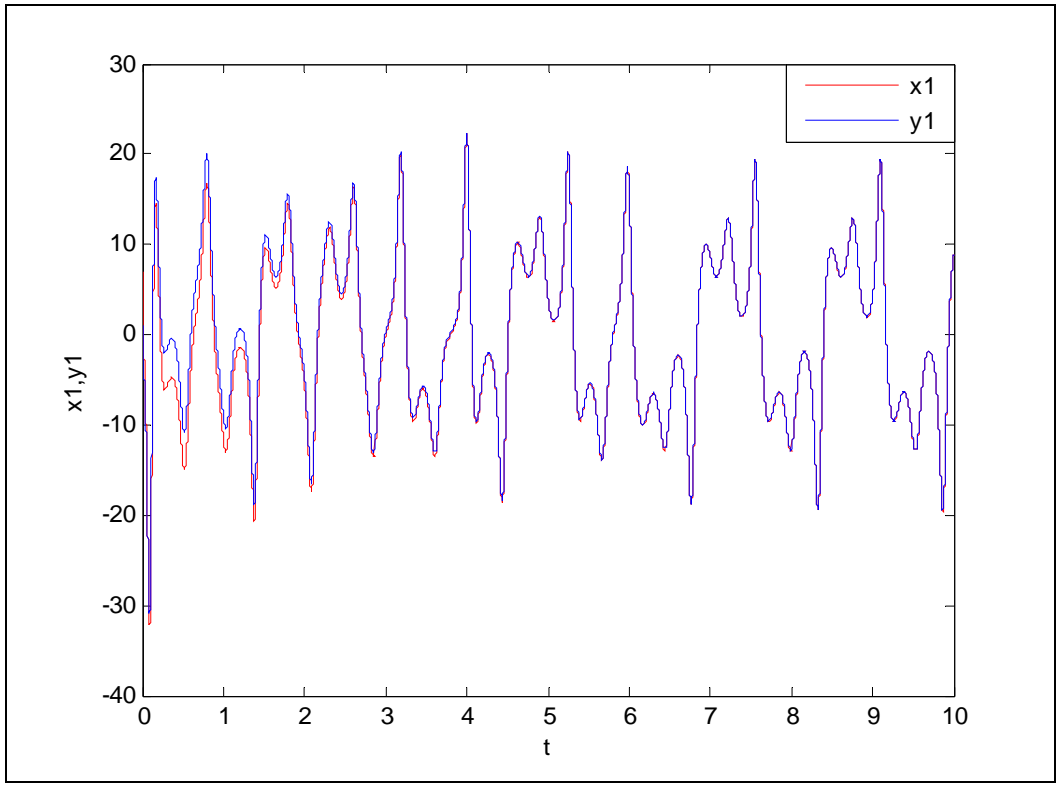

Fig. 1(a). The state trajectories of the system (25) and system (28) with the sliding mode control method. (signals $x_1$; $y_1$).

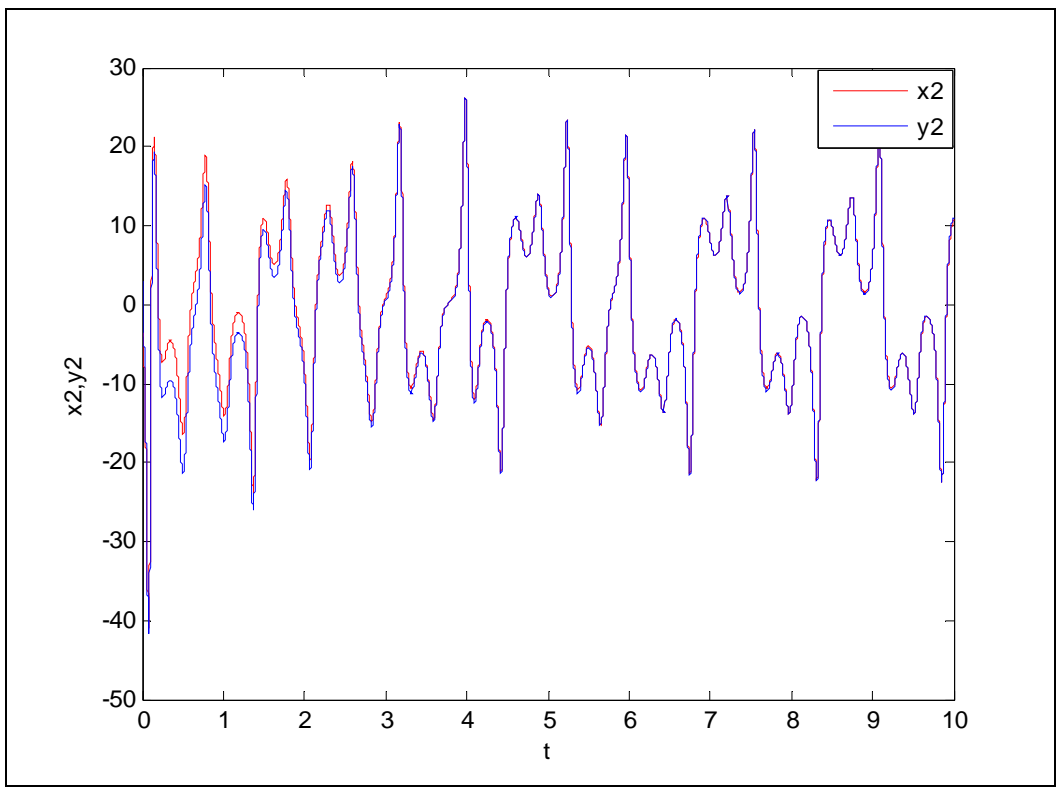

Fig. 1(b). The state trajectories of the system (25) and system (28) with the sliding mode control method. (signals $x_2$; $y_2$).

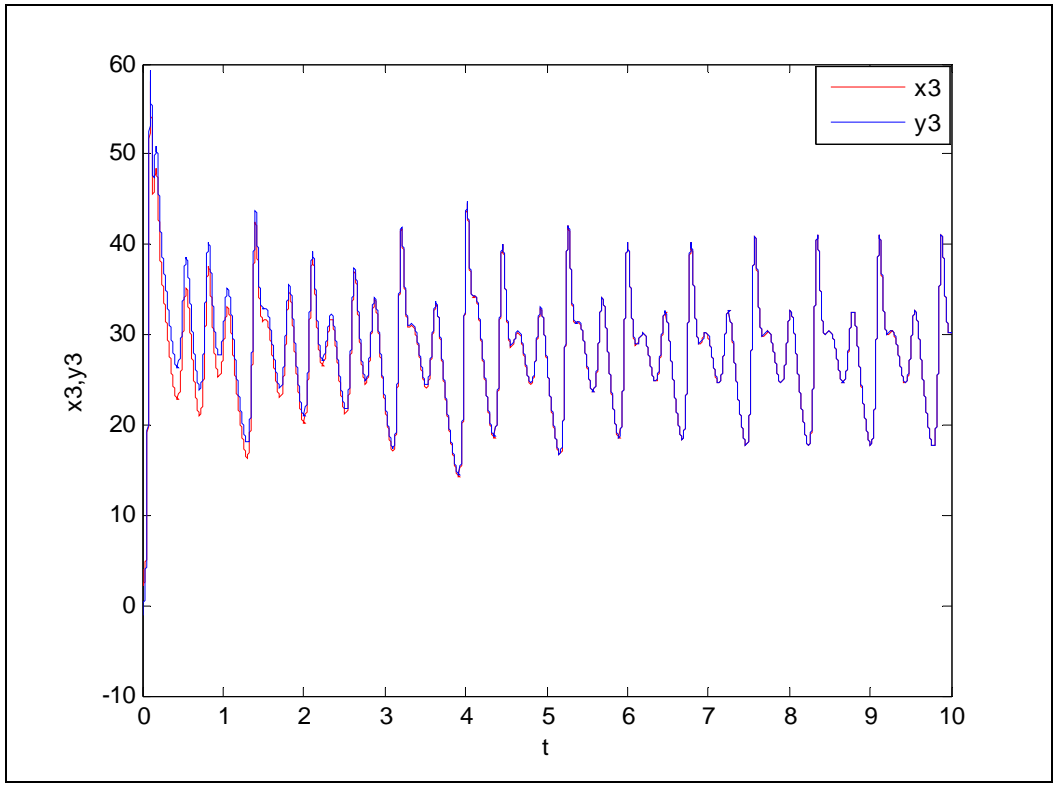

Fig. 1(c). The state trajectories of the system (25) and system (28) with the sliding mode control method. (signals $x_3$; $y_3$).

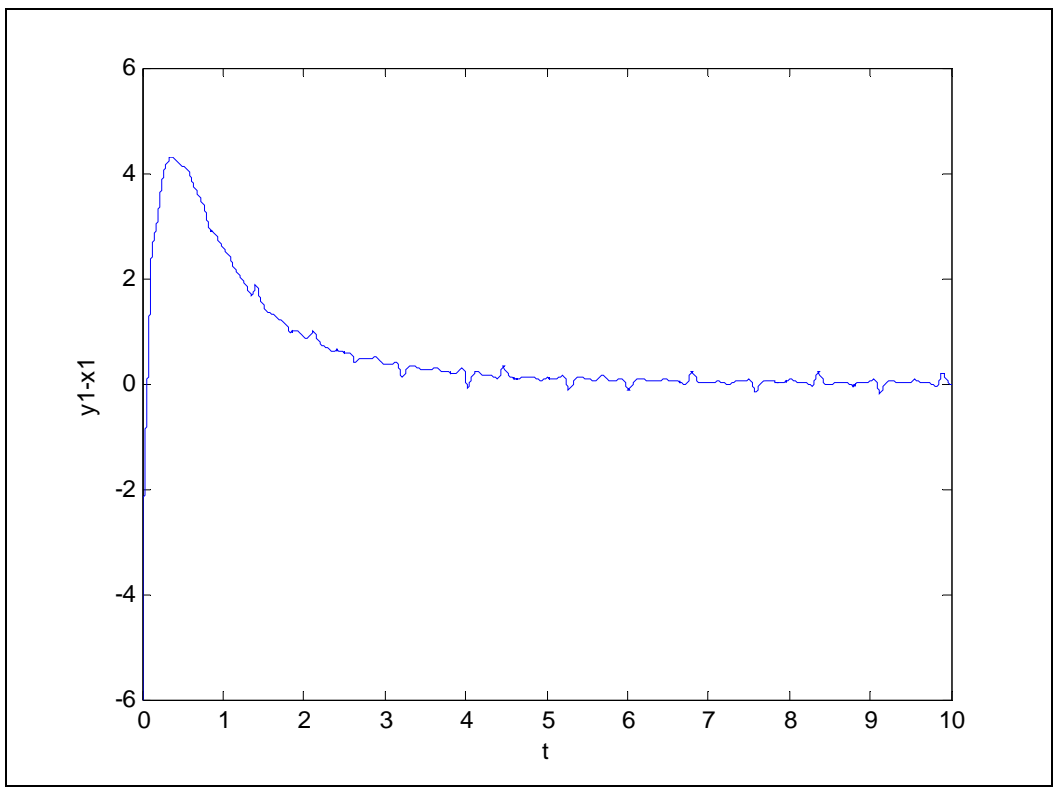

Fig. 2(a). The time evolution of synchronization error $e_1$ of the drive system (25) and response system (28).

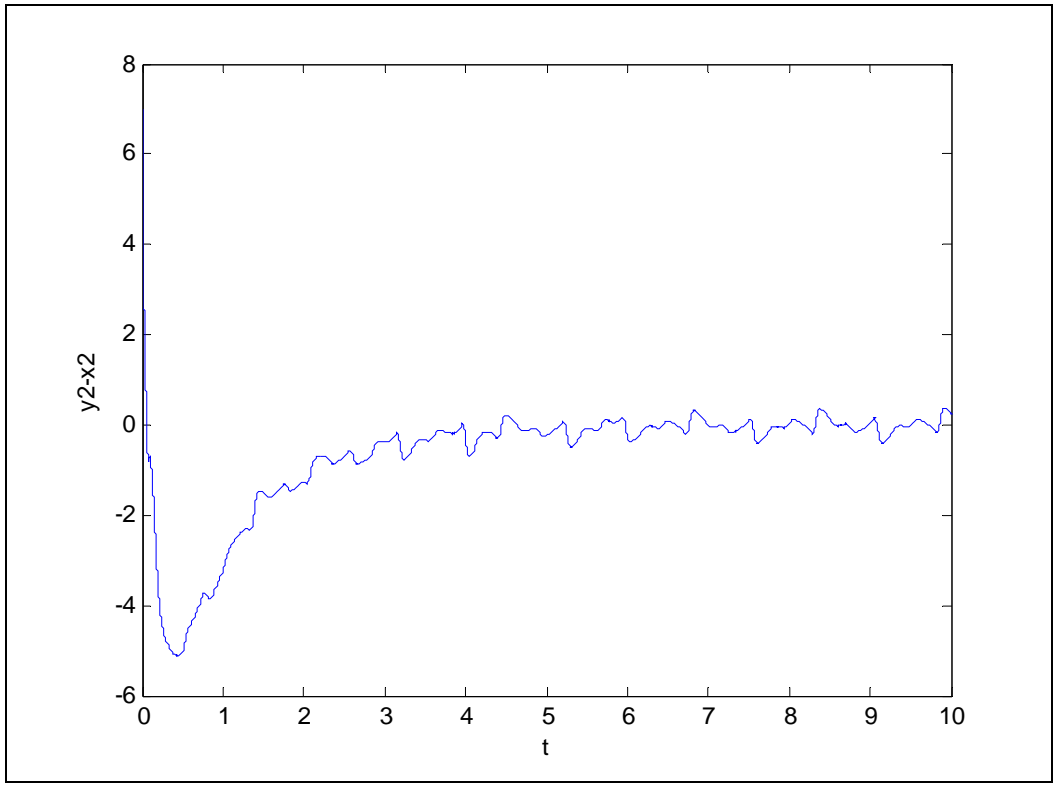

Fig. 2(b). The time evolution of synchronization error $e_2$ of the drive system (25) and response system (28).

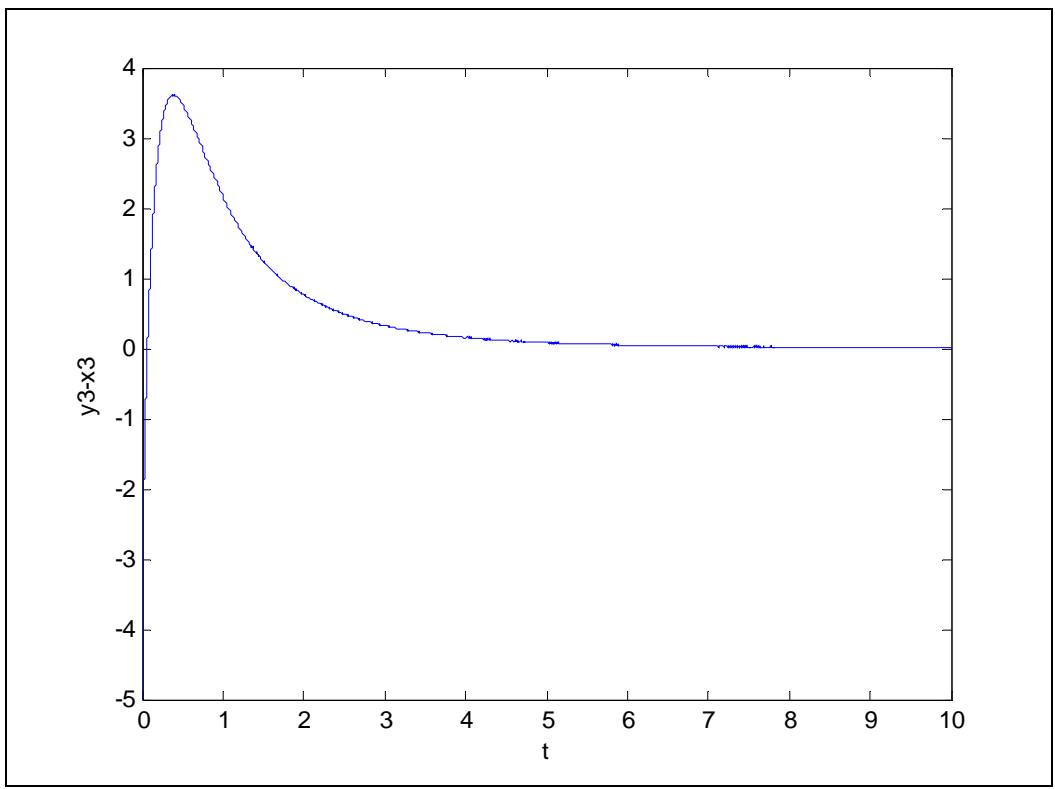

Fig. 2(c). The time evolution of synchronization error $e_3$ of the drive system (25) and response system (28).

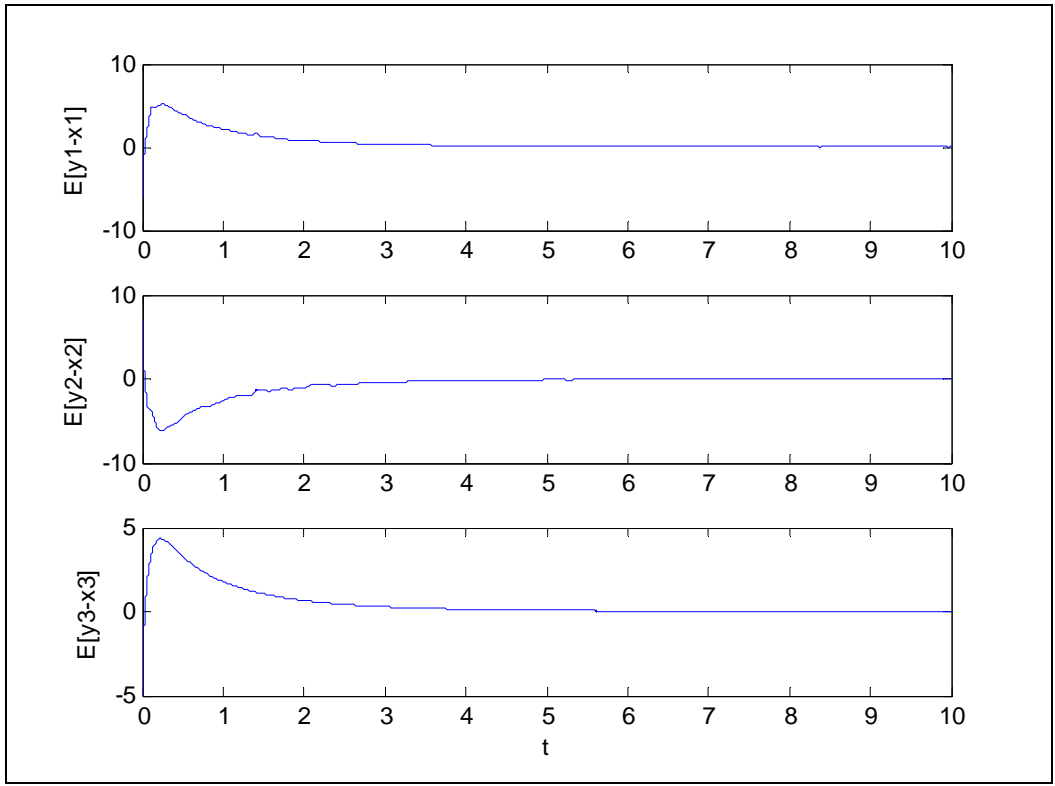

Fig.3. Mean value of synchronization errors $e_1, e_2, e_3$ between system (25) and system (28) using sliding mode method.

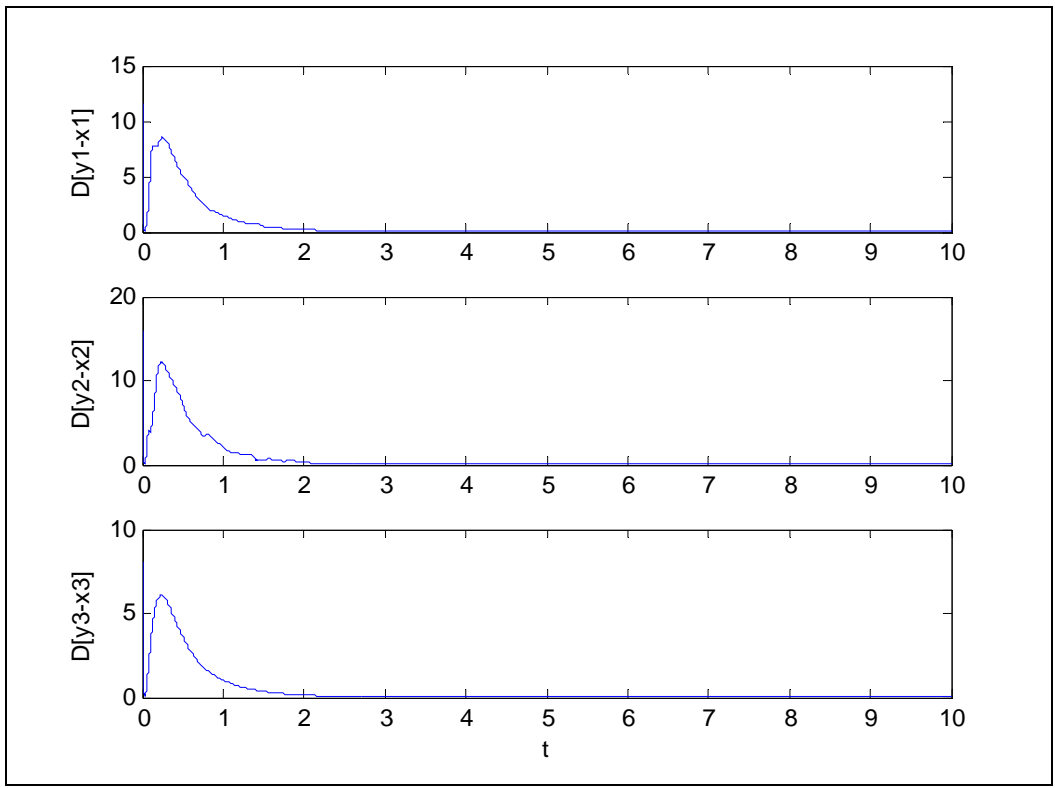

Fig.4. Variance of the state estimation errors $e_1, e_2, e_3$ of the system (25) and system (28) using the sliding mode method.

## 5. Conclusions

In this paper, a new method for synchronizing two different fractional-order chaotic systems is introduced, and the driving chaotic system is excited by Gaussian white noise. A fractional sliding surface is introduced, and the sliding mode controller is proposed for synchronization. Furthermore, convergence property has been analyzed for the error dynamics after adding proposed controllers. It has been shown that by proper choice of the control parameters ($r$ and $\rho$), the fractional-order chaotic systems in uncertain environment are synchronized. Finally, to further illustrate the effectiveness of the proposed controllers, one applies the presented algorithm to the fractional-order Chen and fractional-order Lü systems through numerical simulations. From the simulation results, it is obvious that a satisfying control performance can be achieved by using the proposed method.

## Acknowledgements

This work was supported by the NSF of China (Grant Nos. 11372247, 11102157), Program for NCET, the Shaanxi Project for Young New Star in Science &Technology, NPU Foundation for Fundamental Research and SRF for ROCS, SEM.